\documentclass[usenatbib,useAMS]{mn2e} \usepackage{psfig,graphicx,longtable,times}

\title[Stellar jitter from variable gravitational redshift] {Stellar jitter from variable gravitational redshift: implications for RV confirmation of habitable exoplanets}

\author[Cegla et. al\\]
{H.\,M.\ Cegla$^{1,2}$,  C.\,A.\ Watson$^1$\thanks{E-mail: c.a.watson@qub.ac.uk}, T.\,R.\ Marsh$^3$, S.\ Shelyag$^1$, V.\ Moulds$^1$, S.\ Littlefair$^4$, \newauthor M.\ Mathioudakis$^1$, D.\ Pollacco$^1$, X. Bonfils$^5$\\
$^1$ Astrophysics Research Centre, School of Mathematics \& Physics, Queen's University,
University Road, Belfast BT7 1NN, UK \\
$^2$ Department of Physics \& Astronomy, Vanderbilt University, Nashville, Tennessee 37235, USA\\
$^3$ Department of Physics, University of Warwick, Gibbet Hill Road, Coventry, CV4 7AL\\
$^4$ Department of Physics and Astronomy, University of Sheffield, Sheffield S3 7RH\\
$^5$ UniversitŽ J. Fourier (Grenoble 1)/CNRS, Laboratoire dÕAstrophysique de Grenoble (LAOG, UMR5571), France\\}

\date{\center{\Large Submitted for publication in the Monthly Notices
    of the Royal Astronomical Society \\
\vspace{.5cm} \today}}

\begin{document}

\date{Accepted 2011 December 6.  Received 2011 December 1; in original form 2011 November 17}


\maketitle

\begin{abstract}
A variation of gravitational redshift, arising from stellar radius fluctuations, will introduce astrophysical noise into radial velocity measurements by shifting the centroid of the observed spectral lines. Shifting the centroid does not necessarily introduce line asymmetries. This is fundamentally different from other types of stellar jitter so far identified, which do result from line asymmetries. Furthermore, only a very small change in stellar radius, $\sim$ 0.01\%, is necessary to generate a gravitational redshift variation large enough to mask or mimic an Earth-twin. We explore possible mechanisms for stellar radius fluctuations in low-mass stars. Convective inhibition due to varying magnetic field strengths and the Wilson depression of starspots are both found to induce substantial gravitational redshift variations. Finally, we investigate a possible method for monitoring/correcting this newly identified potential source of jitter and comment on its impact for future exoplanet searches. 
\end{abstract}


\section{Introduction}
\label{sec:intro}
Space-based, photometric surveys have moved us into a new era of exoplanet discovery. The Kepler mission alone has identified 1235 planetary candidates, 54 of which lie within the habitable zone (\citealt{borucki11}). However, radial velocity (RV) follow up is mandatory for planet confirmation. To do this for low-mass planets typically requires cm s$^{-1}$ RV precision.

However, astrophysical noise sources (or stellar jitter) due to spots, plages, granulation and stellar oscillations, for example, become an issue at the m s$^{-1}$ level. These phenomena alter the shape of the stellar absorption lines, injecting spurious or systematic RV signals that may mask or mimic planetary signals. There are numerous examples of this occurring in the literature, e.g. HD 166435, HD 219542, TW Hya, LkCa19 and BD +20 1790 (\citealt{queloz01}; \citealt{desidera04}; \citealt{huelamo08}; \citealt{huerta08}; \citealt{figueria10}). 

Spots and plages can induce RV signals of 1-100 m s$^{-1}$ (\citealt{saar97}). There have been many attempts to remove this large-scale jitter using spot simulations, Keplerian models, bisector analysis, harmonic decomposition, and Fourier analysis etc. (see \citealt{saar97}; \citealt{hatzes02}; \citealt{desort07}; \citealt{dumusque11b}; \citealt{boisse11}; \citealt{melo07}; \citealt{bonfils07}; \citealt{boisse09}; \citealt{queloz09}; \citealt{hatzes10}). Even with these removal techniques, active stars are still not ideal candidates for RV follow-up and are thus often left out of planet surveys. 

Unfortunately, even `quiet' stars (those with little or no activity) exhibit jitter due to granulation and stellar oscillations. Granulation and stellar oscillations can introduce typical RV signals of a m s$^{-1}$ or higher (e.g. \citealt{schrijver00}). For this small-scale jitter, the currently implemented removal technique rests on adapting observational strategies to average out such noise (see \citealt{pepe2011}; \citealt{dumusque11a}). However, this technique is observationally intensive and does not provide information on the nature of jitter. 

In order to confirm low-mass planets around low-mass stars it is of utmost importance that we identify and understand stellar jitter, in all its forms. In this paper, we outline another possible source of stellar jitter due to variations in gravitational redshift. These variations could potentially arise from fluctuations in the stellar radius or changes in the line formation height. 

In Section~\ref{sec:gr}, we discuss the implications that variable gravitational redshift has on exoplanet confirmation/detection. We investigate measurements of solar radius fluctuations  in Section~\ref{sec:radnow}, and in Section~\ref{sec:gr_var} potential mechanisms to produce stellar radius changes. In Section~\ref{sec:disc}, we discuss our findings and outline a possible solution.

\section{Gravitational Redshift}
\label{sec:gr}
As photons escape from the deep gravitational potential well of their star they lose energy, causing a redshift in the observed spectral lines (e.g. \citealt{tayler}; \citealt{lindegren03}). Any variation in the gravitational redshift (GR) would result in a wholesale shift in the stellar absorption line positions which could mask or mimic a planetary signal. Unlike activity induced jitter, however, there are no resultant line asymmetries. Thus, jitter from GR variations may prove more difficult to distinguish from the Doppler-reflex motion caused by a planet. 

Here, we estimate the magnitude of the potential RV signals introduced by such a GR variation. The observed velocity shift in the line centres, for a photon detected at an infinitely large distance from the stellar surface, is 
\begin{equation}
\label{eq:vg}
{V_{grav}} = \frac{GM}{Rc},
\end{equation}
where G is the gravitational constant, M is the mass of the star, R is the radius of the star, and c is the speed of light. Given a fractional stellar radius change (at constant mass), ${\delta R}$, the observed variation in velocity, ${\delta V_{grav}}$, would be
\begin{equation}
\label{eq:deltavg}
{\delta V_{grav}} = \frac{\delta R}{ R} V_{grav}.
\end{equation}
For the case of the Sun, ${V_{grav}}$ is 636 m s$^{-1}$. If, however, the solar radius varied by only 0.01\%, a distant observer would perceive a shift of $\sim$ 6 cm s$^{-1}$. The RV signal produced from the Earth around our Sun is only $\sim$ 9 cm s$^{-1}$. Thus a stellar radius change of $\ge$ 0.01\%, by any means, could potentially mimic or mask a RV signal from an Earth-like planet.  

\begin{figure}
\includegraphics[width=8.5cm]{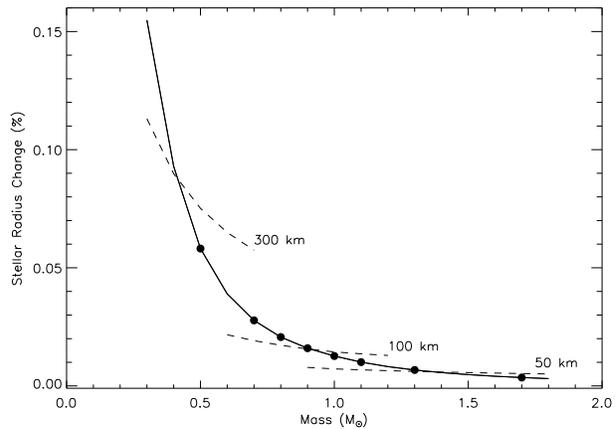}
\caption{Smooth curve represents the stellar radius change (expressed as a percentage) required to induce a ${\delta V{_{grav}}}$ equivalent to an Earth-twin RV signal. Circles represent, from right to left, spectral types: F0, F5, G0, G2, G5, K0, K5, and M0. Dashed curves represent stellar radius variations of 50, 100 and 300 km.} \label{fig:deltaR}
\end{figure}

We are interested in the impact of GR jitter on all habitable exoplanet detections. The habitable zone scales with host star mass, as does the planetary RV signal. The RV signal from a habitable Earth-mass planet around spectral types F0-M0 varies from about 3 - 50 cm s$^{-1}$. This means a much larger change in stellar radius is required for lower-mass host stars to produce a ${\delta V_{grav}}$ large enough to mask or mimic a planetary RV signal. This is shown in Figure~\ref{fig:deltaR}, which displays the stellar radius change (expressed as a percentage) required to cause a ${\delta V_{grav}}$ equal in magnitude to the RV signal due to a habitable Earth-mass planet orbiting stars of masses 0.3 - 1.8 $M_{\odot}$.

From Figure~\ref{fig:deltaR} we find, for M dwarfs, a relatively large stellar radius fluctuation ($\sim$ 0.1\%) would be necessary to generate a substantial enough GR variation to mask or mimic a habitable Earth-mass planet. Additionally, because a habitable planet orbiting a M dwarf has a period of days or weeks this radius change would need to occur over this timescale as well. For G2V and higher mass stars, we see a small ($\sim$ 0.01\%) change in stellar radius, over a year or more, has the potential to induce GR jitter capable of masking or mimicking an Earth-like RV signal. Thus, GR variations are unlikely to pose a problem when planet hunting around M dwarfs due to the relatively large stellar radius change required over relatively short timescales. However, GR jitter could pose more significant issues when searching around solar-like stars on account of the smaller GR variation required over far longer timescales.

Because a radius change means that the stellar surface is receding or approaching the observer, we have also examined the Doppler-shift imparted by this motion. Equating the amplitude of the RV signal due to a change in stellar radius (assumed to be a uniform contraction/expansion rate) over time to the corresponding $V_{grav}$ gives the characteristic timescale, $t_{grav}$, at which GR variations begin to dominate,
\begin{equation}
\label{eq:tgr}
t_{grav} > \frac{R^2c}{GM}.
\end{equation}
Thus, for low-mass stars we can rule out any radius fluctuations, such as stellar oscillations (e.g. the 5 minute solar oscillation), that occur on a timescale less than about 10 days as sources of GR jitter. 

\section{The Solar Radius}
\label{sec:radnow}
Given that stellar radius changes  can potentially mask or mimic an Earth-twin, we have searched for evidence of solar radius changes in the literature. Theoretically, there are a number of predictions of solar radius fluctuations ranging from $\sim$ 0.001 to 0.01\% (where GR jitter becomes an issue), largely due to magnetic field interplay. For example, \cite{stothers06} argues using the virial theorem that energy conservation requires a stellar radius decrease when magnetic activity increases. \cite{fazel08} suggest that magnetic pressure interferes with convection and causes the size of solar granules to decrease, leading to an apparent decrease in the solar radius. \cite{mullan07} also propose that magnetic pressure could contribute to hydrostatic support within the star, again leading to stellar radius fluctuations. 

Table~\ref{tab:rad} provides a summary of the most recent solar radius variation measurements that we are aware of (this table originates from \cite{thuillier05}; more recent results have been appended). While some authors report solar radius changes of 0.01-0.1\%, the most recent space-based observations indicate that the solar radius is stable to within $\sim$ 0.0001\%. It is believed that the discrepancies between the reported solar radius changes may be due to atmospheric contamination of the ground-based observations. However, there is still an unexplained discrepancy between results from stratospheric balloons and those made from space (\citealt{bush10}). In conclusion, there is no concrete observational evidence of appreciable solar radius changes on the timescales comparable to the solar activity cycle. 

\section{Possible Sources of GR Variations}
\label{sec:gr_var}
Given the lack of any clear observational evidence for solar radius changes of the magnitude required to raise substantial GR jitter, we now look at other possible mechanisms that may be important for other stars/spectral types. We note that each of these mechanisms will have their own jitter, however, it is beyond the scope of this letter to quantify these effects and we only consider the GR component.

\subsection{Magnetic Activity} 
 
\subsubsection{The Applegate Effect}
\label{sec:app}
Quasi-periodic orbital period modulations, over a timespan of a few years, have been observed in several eclipsing binary systems (e.g. \citealt{baptista03}; \citealt{borges08} and references therein); these are believed to be caused by the Applegate effect (\citealt{applegate92}). According to the Applegate effect the stellar gravitational quadrupole moment changes due to the magnetic activity cycle, redistributing angular momentum within the star's interior. The redistribution of angular momentum alters the oblateness of the star, thereby also causing stellar radius fluctuations. We attempt to estimate the magnitude of this stellar radius change, first in the close binaries from \cite{lanza99} (excluding evolved stars) and then later extrapolate to exoplanet host stars.  

According to \cite{rudiger02}, we can use the orbital period modulations, $\delta P / P$, observed in binaries to estimate the change in gravitational quadrupole moment, $\delta{J{_2}}$, via
\begin{equation}
\frac{\delta P}{P} = -3 \frac{R^2}{a^2} \delta{J{_2}},
\label{eqn:pp}
\end{equation}
where a is the semi-major axis.
The gravitational quadrupole moment is also related to oblateness, ${f}$ (\citealt{dicke70}), and the stellar radius (\citealt{barnes03}) via
\begin{equation}
\label{eq:j2}
J{_2} \approx \frac{2}{3}f \approx \frac{2}{3} \frac{R_{eq}-R_{p}}{R_{eq}},
\end{equation}
where R$_{eq}$ and R$_{p}$ are the stellar equatorial and polar radius, respectively. Differentiating equation \ref{eq:j2} with respect to R$_{eq}$ (where we assume R$_{eq}$ $\approx$ R$_{p}$) gives,
\begin{equation}
\delta{J_{2}} \approx \frac{2}{3} \frac{\delta R}{ R}.
\end{equation}
Thus, we can estimate the fractional stellar radius change from the observed orbital period variation,
\begin{equation}
\label{eq:rrpp}
\frac{\delta R}{R} \approx -\frac{a^2}{2R^2} \frac{\delta P}{ P}.
\end{equation}

Using equation \ref{eq:rrpp}, and the binary star parameters from \cite{lanza99}, we estimate stellar radius changes of 0.001-0.1\% for the rapidly rotating low-mass stars in these systems. Unfortunately, such measurements of orbital period variations do not exist for stars with rotation rates comparable to those of exoplanet host stars (which are generally slowly rotating). We therefore estimate the potential stellar radius change for these stars assuming that a fixed energy budget is available to drive the Applegate mechanism (as done by \citealt{watson10}). According to \cite{watson10}, the change in energy, $\delta E$, required for such orbital period modulations can be described as,

\begin{equation}
\label{eq:energy}
\delta E = \frac{1}{6} \frac{G^2M^4a^4}{\Omega^2R^8M_s}\Big(\frac{\delta P}{P}\Big)^2,
\end{equation}
where $\Omega$ is the is the stellar rotational angular velocity and M$_s$ is the mass of the convective shell.

As in \cite{watson10}, the available energy budget is constrained to 10\% of the stellar luminosity such that, 
\begin{equation}
\label{eq:energybug}
\delta E < \frac{\alpha LT}{\pi},
\end{equation}
where L is luminosity, T is the timescale of the gravitational quadrupole moment changes, and $\alpha$ is 0.1. Substituting equation \ref{eq:rrpp} and \ref{eq:energybug} into equation \ref{eq:energy}, we find that the fractional change in stellar radius for a low-mass star over a year, due to the Applegate mechanism, contributes less than a cm s$^{-1}$ of GR jitter. Therefore GR jitter from the Applegate effect is most likely negligible in slowly rotating, low-mass exoplanet host stars. 
 
\subsubsection{Inhibition of Convection}
\label{sec:conv}
The presence of a magnetic field has long been thought to inhibit convection (\citealt{makita58}; \citealt{spruit00}). A decrease in the efficiency of convection is thought to result in the suppression of hot up-rising granules, possibly leading to an effective decrease in the observed stellar radius as the magnetic field is increased. Since the magnetic field over an activity cycle is variable, this mechanism may cause apparent stellar radius changes on the timescale of years. 

To further explore the impact that convective inhibition has on the effective stellar radius, we have simulated solar convection using the MURam code (\citealt{vogler05}). The code solves large-eddy radiative three-dimensional magnetohydrodynamic equations on a Cartesian grid, and employs a fourth-order central difference scheme to calculate spatial derivatives. The numerical domain has a physical size of 12 x 12 Mm$^{2}$ in the horizontal direction and  1.4 Mm in the vertical direction (see \citealt{shelyag11} for details). Continuum contribution functions were calculated in a narrow ($1~\mathrm{nm}$) band centred at $630.2~\mathrm{nm}$ using the method described in \citet{jess11}. Since the total mass in the numerical domain is conserved and the gravitational acceleration is assumed constant throughout the domain, the change in height of the photospheric radiation formation in the model can be directly related to the change in the solar radius. 

Models of the solar photosphere were generated assuming an average magnetic field of 50, 100, 150 and 400 G. To reduce differences introduced by stellar oscillations the contribution functions for each model were calculated at the same oscillation phase. We find that the 50 G model has a centre of gravity $\sim$ 3 km higher than that of the 100 G, $\sim$ 7 km higher than the 150 G, and $\sim$ 40 km higher than the 400 G model. This corresponds to a GR jitter ranging from $\sim$ 0.3 - 4 cm s$^{-1}$. Therefore, our simulations suggest that the inhibition of convection due to increasing magnetic field strength can be a possible source of GR jitter.

\subsection{Wilson Depression of Starspots}
\label{sec:spots}
It is well known that sunspots are depressed compared to the photosphere (the Wilson depression). The depth of the Wilson depression has been estimated in many previous studies, ranging from 500 - 2500 km (e.g \citealt{suzuki67}; \citealt{balthasar83}; \citealt{watsonf09}). The spectral contribution of a sunspot with a Wilson depression of 1400 km would be gravitationally red-shifted by $\sim$ 1.3 m s$^{-1}$ compared to the immaculate photosphere. However, as a spot contributes only a small amount of the total observed stellar flux, the net effect on the RVs will be much smaller. 

In order to estimate the amplitude of GR jitter from the Wilson depression, we have created a model spotty star. Line-lists for spectral types G2V and K5V were downloaded from the Vienna Atomic Line Database (VALD--see \citealt{kupka00}; \citealt{kupka99}). Synthetic G2V and K5V spectra were then generated by convolving these line-lists with an intrinsic line profile with macroturbulence calculated from \cite{valenti05}, $v$sin$i$ = 0 km s$^{-1}$, an instrumental broadening = 2.6 km s$^{-1}$, and microturbulence = 1.5 km s$^{-1}$. The total synthetic spectrum for the spotty star was then modelled as the sum of the G2V spectrum (representing the immaculate photosphere) and the K5V spectrum (representing the cooler spots); we assumed a spot filling factor of 0.01. We tested two different depths for the Wilson depression, $\sim$ 1400 km and 3500 km; to do this the spectral contribution from the spots were shifted by $\sim$ 1.3 and 3.2 m s$^{-1}$, respectively (the shift expected due to GR). To measure the net RV shift induced by the GR component of the Wilson depression, we cross-correlated the non-shifted model spectrum with each of the two GR-shifted model spectrums. This process was repeated assuming a  cooler spot, created using a M0V spectrum. 

For the model star with K5V spots, we find net RV shifts of $\sim$ 2 and 3 cm s$^{-1}$ for Wilson depression depths of $\sim$ 1400 km and 3500 km, respectively; for the model star with M0V spots we find net RV shifts of $\sim$ 1 and 2 cm s$^{-1}$. The model star with K5V spots produces larger net RV shifts because the spot component is brighter and therefore contributes more to the overall flux. The depth of the Wilson depression determines the magnitude of the RV shifts as a larger depth induces more GR jitter. Overall, for a quiet solar-type star, GR jitter from the Wilson depression is most likely a few cm s$^{-1}$, which is not negligible when searching for habitable low-mass planets. We stress, however, that the presence of spots will introduce other sources of jitter. For example, since convection is inhibited within a spot, the convective blueshift (typically 100's m s$^{-1}$, \citealt{asplund00}) is also inhibited. This would significantly shift the spotty spectrum relative to the immaculate photosphere.

\section{Discussion}
\label{sec:disc}

Changes in the observed spectral lines have the potential to mask or mimic the presence of a planet. In this paper, we have shown that a stellar radius fluctuation of even 0.01\% can induce a GR variation large enough to shift the centroid of spectral lines by $\sim$ 6 cm s$^{-1}$. We have also calculated a characteristic timescale, $t_{grav}$, over which stellar radius fluctuations must occur for GR jitter to dominant the RVs. From this we have found that GR jitter is a dominate noise source for stellar radius changes occurring on periods longer than $\sim$ 10 days for low-mass stars. In light of this, we explored current solar radii measurements and potential mechanisms to produce such stellar radius changes.

We find there is no universally accepted answer to the constancy of the solar radius. Most recent space-based measurements indicate a solar radius change of only 0.0001\%. However, this is not in agreement with measurements from ground-based instruments, though there is reason to believe these may be systematically affected by the atmosphere. Missions such as PICARD and SDO may hold the key to high precision solar radius measurements in the future (\citealt{picard06}; \citealt{pesnell11}).

We explored the origin of potential stellar radius changes. For slowly rotating, low-mass stars we find that the Applegate effect is most likely negligible. However, from our simulations we find changes in surface convection patterns, due to varying magnetic field strengths, alter the effective stellar radius sufficiently for appreciable GR jitter in the habitable planet regime. In addition, the Wilson depression of starspots was also found to contribute a substantial GR shift when compared to Earth-like RV signals. Therefore GR jitter could be an important and hitherto unrecognised astrophysical noise source. We note here that although significant, GR jitter may be overwhelmed by other line-altering mechanisms such as the Paschen-Back effect, which will be discussed in a forthcoming paper. 
 
Nonetheless, confronted with this new noise source, we believe precise measurements of transit duration variations (TDVs) could be the key to removing GR jitter. As the stellar radius expands or contracts, the transit duration would increase or decrease accordingly. Measurements of the TDV would allow us to calculate the change in stellar radius and thus monitor the GR variation at some level.

This method does have its limitations as very precise photometric measurements are necessary to detect the potential TDV. For example, if the solar radius were to change by 60 km ($\sim$ 0.01\% R), then the transit duration of the Earth would differ by $<$ 3 s. The lower limit of the TDV is set by transits with an impact factor of zero, while grazing planets would exhibit larger TDVs, and transits may even disappear in extreme cases. Naturally, however, this would not enable the correction of GR jitter due to depressed spots.  

In conclusion, we have identified a new potential source of stellar jitter in the form of GR variations and outlined different mechanisms capable of generating cm s$^{-1}$ variations. This may be important in the RV follow-up and confirmation of low-mass terrestrial planets and Earth-like worlds.

\section*{\sc Acknowledgments}
This research has made use of NASA's Astrophysics Data System Bibliographic Services. HMC acknowledges support from a Queen's University Belfast university scholarship. CAW would like to acknowledge support by STFC grant ST/I001123/1. TRM was supported by an STFC Rolling Grant during the course of this work. SPL is supported by an RCUK fellowship.

\bibliographystyle{mn2e}
\bibliography{abbrev,refs}

\onecolumn
\begin{table}
\caption{Summary of observational measurements of solar radius fluctuations over an activity cycle, published primarily in the last ten years. Results are separated by location: G: ground-based, A: atmosphere-based, S: space-based.}
\centering
\begin{tabular}{ccccccc}
    \hline
    \hline
     References ¥ & Location ¥ & Method/ ¥ & $\lambda$ ¥ & Period  ¥ & Radius ¥ & ¥ \\ 
    ¥ & ¥ & Instrument ¥ & ¥ & ¥ & ¥ ¥ ¥ ¥ ¥ ¥ Change (\%) ¥ & ¥ \\ 
    \hline  
    \cite{selhorst04} ¥ & G ¥ & ¥  Radio Waves & 16 GHz ¥ & 1992-2003 ¥ & 0.1 ¥  \\ 
    \cite{sofia85} ¥ & G ¥ & Eclipses ¥ & Visible ¥ & 1925/1979 ¥ & 0.01 ¥ \\  
    \cite{sveshnikov02} ¥ & G ¥ & Mercury ¥ & Visible ¥ & 1631-1973 ¥ & 0.01 ¥ \\  
    \cite{delmas02} ¥ & G ¥ & Astrolabe ¥ & 540 nm ¥ & 1978-2004 ¥ & 0.01 ¥ \\
    \cite{reisneto03} ¥ & G ¥ & Astrolabe ¥ & 563.5 nm ¥ & 1998-2000 & 0.01 ¥ \\ 
    \cite{noel04} ¥ & G ¥ & Astrolabe ¥ & 540 nm ¥ & 1991-2002 ¥ & 0.01 ¥ \\  
    \cite{emilio05} ¥ & G ¥ & Astrolabe ¥ &  ¥ & 1972-1998 & 0.01 ¥ \\ 
    \cite{chapman08} ¥ & G ¥ & Scanning ¥ & 672.3 nm ¥ & 1986-2004 ¥ & 0.01 ¥ \\ 
    \cite{penna10} ¥ & G ¥ & Astrolabe ¥ & ¥ & 1997-2008 ¥ & 0.01 ¥ \\   
    \cite{brown98} ¥ & G ¥ & Meridian ¥ & 800 nm ¥ & 1981-1987 ¥ &$<$ 0.01 ¥ \\ 
    \cite{wittmann03} ¥ & G ¥ & Meridian ¥ & 585 nm ¥ & 1972-2002 ¥ &$<$ 0.01 ¥ \\
    \cite{lefebvre06} ¥ & G ¥ & Scanning ¥ & 525 nm ¥ & 1970-2003 ¥ &$<$ 0.01 ¥ \\  
    \cite{badache06} ¥ & G ¥ & Astrolabe ¥ & ¥ & 1998-2003 ¥ & 0.001-0.01 ¥ \\ 
    \cite{kilic05} ¥ & G ¥ & Astrolabe ¥ & 550 nm ¥ & 2001-2003 ¥ & 0.001 ¥ \\  
    \cite{egidi06} ¥ & A ¥ & SDS ¥ & 600 nm ¥ & 1992-1996 ¥ & 0.01 ¥ \\ 
    \cite{bush10} ¥ & S ¥ & MDI ¥ & 676.8 nm ¥ & 1995-2009 ¥ &$<$ 0.001 ¥ \\     
    \cite{antia03} ¥ & S ¥ & Helioseis ¥ & ¥ & 1995-2004 ¥ & 0.0001 ¥ \\ 
    \cite{dziembowski05} ¥ & S ¥ & MDI ¥ & ¥ & 1996-2004 ¥ & 0.0001 ¥ \\  
  \hline
  \end{tabular}
  \label{tab:rad} 
  \end{table}

\twocolumn

\end{document}